\pdfoutput=1
\documentclass[%
 amsmath,amssymb,
 prx,
 lengthcheck,%
]{revtex4-1}

\usepackage{graphicx}
\usepackage{dcolumn}
\usepackage{bm}
\usepackage{hyperref}

\def\eg{{\it e.g.\ }} 
 
\def\p{\partial}
\newcommand{\bV}{{\bf V}}

\newcommand{\divergence}{{\rm div}}
\newcommand{\odivergence}{{\rm odiv}}
\newcommand{\grad}{{\rm grad}}

\def\ubar{\bar{u}}

\def\rhobar{\bar{\rho}}
\def\mbar{\overline{\rho u}}

\def\xb{\mathbf{x}}

\newtheorem{theorem}{Theorem}[section]

\newtheorem{definition}{Definition}[section]

\newcommand{\eref}[1]{(\ref{#1})}

\def\drawline#1#2{\raise 2.5pt\vbox{\hrule width #1pt height #2pt}}

\begin{document}

\preprint{APS/123-QED}

\title{Observable Divergence Theorem: \\Evolution Equations for Inviscid Regularization of Shocks and Turbulence\vspace{-2mm}}\vspace{-2mm} 
\thanks{The research in this paper was partially supported by the AFOSR.}

\author{Kamran Mohseni}
\email{mohseni@colorado.edu}
 \homepage{http://enstrophy.colorado.edu/~mohseni/}
 \altaffiliation{Department of Aerospace Engineering Sciences, University of Colorado, Boulder, CO}


%



\begin{abstract}
The divergence theorem of Gauss plays a central role in the derivation of the governing differential equations in fluid dynamics, electrodynamics, gravitational fields, and optics. Many of these phenomena, particularly shocks and turbulence in fluids, are multi-scale in nature and prone to continuous generation of high wave-number modes or small scales. In practical applications, however, one is interested in an evolution equation for the large scale quantities without resolving the details of the small scales. As a result, there has been a significant effort in developing time-averaged and spatially-filtered equations for large scale dynamics from the fully resolved governing differential equations. One should realize that by starting from these fully-resolved equations (\eg the Euler or Navier-Stokes equations) to derive an averaged evolution equation one has already taken the limit of the wave-numbers approaching infinity with no regards to our observational abilities at such a limit. 
As a result, obtaining the evolution equations for large scale quantities (low wave-numbers) by an averaging or filtering process is done after the fact in order to somehow regularize the irregular behavior at high wave-numbers. This could explain many of the theoretical and computational difficulties with the Euler or Navier-Stokes equations. Here, a rather different approach is proposed. {\it The averaging process in implemented before the derivation of the differential form of the transport equations}. A new {\textbf {\textit {observable}}} divergence concept is defined based on fluxes calculated from observable quantities at a desired averaging scale, $\alpha$. An observable divergence theorem is then proved and applied in the derivation of the observable and regularized transport equations. We further show that the application of the observable divergence theorem to incompressible flows results in a formal derivation of the inviscid Leray turbulence model first proposed in 1934. Finally, some numerical simulations are presented for the observable inviscid compressible Euler equations with shocks.  It is argued that such a methodology in deriving fluid evolution equations removes many of the theoretical and computational difficulties in multi-scale problems such as turbulence and shocks.  

\begin{description}
\item[PACS numbers] {47.40.Nm, 47.27.-i, 47.35.-i, 47.40.-x}\vspace{-4mm}
\end{description}
\end{abstract}

\pacs{Valid PACS appear here}
\maketitle



\paragraph{Introduction.} For the last several years we have been promoting the idea of simultaneous inviscid regularization of shocks and turbulence in inviscid flows. While shocks and turbulence are often treated as separate problems \footnote{Shocks are treated as discontinuity formation in hyperbolic equations in compressible flows while turbulence is considered as a closure problem which occur in both compressible and incompressible flows.}, they share some similar characteristics. The primary shared feature between them is a {\it high wavenumber cascade} processes created by the nonlinear convective term $\mathbf{u}\cdot \nabla\mathbf{u}$. Traditionally, this high wavenumber irregularity is regularized and controlled by the addition of a viscous term or a random walk in the Euler equations. While successful in describing many complex flow phenomena, this approach is still facing significant theoretical and computational challenges in high wavenumbers (or high Reynolds number flows)\cite{LionsPL:98ab}. In this investigation, we take a different approach and explore a formal derivation of an inviscid regularization of this high wavenumber behavior in shocks and turbulence. The essence of the technique is in the re-derivation of the divergence theorem of Gauss for field quantities that are averaged. The divergence theorem of Gauss in calculus is revisited in order to prove an analogous theorem for a control volume where the fluxes at the faces of the fluid element are calculated based on {\it averaged} flow quantities over a certain length scale as opposed to a fully resolved one.

Our work here was motivated by our earlier results in inviscid regularization of the 1D and multidimensional Burgers equations \cite{Mohseni:08w, Mohseni:09h}
where the nonlinear convective term is replaced by $\overline{\mathbf{u}}\cdot \nabla\mathbf{u}$; $\overline{\mathbf{u}}$ is an averaged velocity. We have shown that under certain conditions the solution to our regularized Burgers equation exists and is unique at all times and it converges to the entropy solution of the Burgers equation. Norgard and Mohseni \cite{Mohseni:10k} has recently extended some of these results to 1D compressible Euler equations. Supporting numerical simulations were presented and it was shown that these equations share the same traveling wave solutions and entropy solutions with the Euler equations. This manuscript provides a formal derivation of our regularized Euler equations from basic principles and its extension to multi-dimensional flows. It should be noted, while a regular viscous term could be included in our analysis without any difficulty that is not necessary for our purposes here. 

\paragraph{Averaging Kernels. }\label{sec:Filter}
In the following sections we are required to calculate the flux of some conserved quantities which depends on an averaging process. A general class of averages for a function $f$ can be defined by the convolution operation as
$
  \overline{f} = g \ast f,
$
where $g$ is the kernel of the convolution. In such an averaging there are several guidelines that seem intuitively reasonable; see \cite{Mohseni:08w, Mohseni:09h}. As listed in Table \ref{FilterProperties}, the averaging kernel should be non-negative, decreasing and radially symmetric. Such a filter $g$ can be associated with a characteristic wavelength $\alpha$. This parameter is introduced by scaling the filter as such,
$  g^{\alpha} = \frac{1}{\alpha} g\left(\frac{\mathbf{x}}{\alpha}\right)$,
or in Fourier domain
$  \widehat{g^{\alpha}}(\mathbf{k}) = \widehat{g} (\alpha \mathbf{k})$.
Thus as $\alpha$ becomes smaller, the wavelength where the filter exerts influence also become smaller. With this scaling, the filter remains normalized, non-negative, decreasing and isotropic. 
Furthermore, as $\alpha \rightarrow 0, \,  g^\alpha$ approaches the Dirac delta function. 
When convolving $g^\alpha$ with $f$, features in $f$ that have length scales less than $\alpha$ will be averaged out.

\begin{table}
\caption{\label{FilterProperties} Requirements for the averaging kernel.}\vspace{-2mm}
\begin{center}
\begin{tabular}{|l|c|}
\hline
Properties & Mathematical Expression\\
\hline
Normalized & $\int g$=1 \\
Nonnegative & $g(\xb) >0,\, \forall \xb $   \\
Decreasing & $|\xb_1|\ge |\xb_2| \Rightarrow g(\xb_1) \le g(\xb_2)$  \\
Symmetric  & $|\xb_1|= |\xb_2| \Rightarrow g(\xb_1) = g(\xb_2)$    \\
\hline
\end{tabular}\vspace{-9mm}
\end{center}
\end{table}

\paragraph{Observability of a Vector Field. }

Our ability to observe and measure (physically or numerically) any fluid property is limited by the spatial resolution of our  experimental or numerical technique. For instance, if one uses hotwire annemometry for measuring turbulence velocity components, the spatial resolution of the measurement is always limited by the size of the wire. In fact, what is actually measured is an averaged value of the velocity over a length scale of the order of the size of the hotwire. 
Similarly in numerical simulations, one is limited by scales of the order of the mesh size. As a result, what we often observe is an averaged quantity rather than the mathematical limit where the observation volume approaches zero. This observation has significant ramifications when one calculates the flux of a quantity at the face of a volume surrounding a point of interest. Before we investigate this, we will revisit the divergence of a vector field and the divergence theorem of Gauss. We will then extend these concepts to the case of a fluid volume with limited observability.

\paragraph{Divergence of a Vector Field. \label{sec:Divergence}}
In writing the differential form of the conservation laws for a fluid, one considers the flux of the conserved quantities from the faces of a fluid volume. These fluxes are then related to the divergence of the conserved quantity inside the volume. 
Let $\Omega$ be a small and closed region centered at point $\mathbf{x_0}$ whose boundary is represented by the orientable surface $S$ and its volume is $\Delta V$. Let $\mathbf{F}$  be a vector function in $C^1(\Omega)$. We denote the flux of the vector field $\mathbf{F}$ through the surface $S$ by $\Phi(\Omega)$. The divergence of the vector field $\mathbf{F}$ around the point $\mathbf{x_0} \in \Omega$  is defined as the limit of the {\it flux density of $\mathbf{F}$ over $\Omega$ around the point $\mathbf{x_0}$}, that is\vspace{-3mm}
\begin{equation}
\!\!\! \divergence \mathbf{F}(\mathbf{x_0}) \stackrel{\text{\tiny def}}{=} \!\! \lim_{\Delta V \rightarrow 0} \dfrac{\Phi(\Omega)}{\Delta V (\Omega)}  = \!\! \lim_{\Delta V \rightarrow 0} \dfrac{1}{\Delta V } \iint\limits_{S(\Omega)} \mathbf{F \cdot n} \; dS\vspace{-3mm}
\label{eq:Divergence2}
\end{equation}
where $\mathbf{n}$ is the outer unit normal vector on $S$. Therefore, the divergence of a vector field is basically the {\it volume averaged} flux of the vector field out of $\Omega$; namely flux per unit volume. Note that by taking $\delta V \rightarrow 0$ we have already made the assumption that the vector field $\mathbf{F}$ is observable with infinite resolution. This assumption will be challenged later in this manuscript. 


A well-known result from the vector calculus is the divergence theorem of Gauss reproduced below \cite{HildebrandFB:62a}. 

\begin{theorem} {Divergence Theorem of Gauss.} Let $\Omega$ be a region with surface boundary $S$ oriented outward. Let $\mathbf{F}(\mathbf{x}) \in C^1(\Omega)$.
Then\vspace{-3mm}
\begin{equation}
  \iiint_\Omega \divergence \mathbf{F} \; dV = \iint_S \mathbf{F \cdot n} \; dS.\vspace{-1mm}
\end{equation}
\end{theorem}

\paragraph{Divergence of an Averaged Vector Field. \label{sec:ODivergence}}
In order to extend the concept of the divergence of a vector field, discussed above, to the case of a finite volume, where the fluxes are calculated based on averaged quantities, we will adopt the definition of the divergence as defined in equation (\ref{eq:Divergence2}); that is a volume-averaged flux. We now follow the standard steps in the proof of the divergence theorem of Gauss with a more careful examination of the flux approximation at the faces of the volume. Before this, we first introduce the following definition.\vspace{-1mm}
\begin{definition}
{Observable Divergence.} Consider a function $f$ and a vector field $\mathbf{V}$. We define the observable divergence of a vector field $\mathbf{F} = f \mathbf{V}$ as\vspace{-1mm}
\begin{equation}
 \odivergence \, \mathbf{F} = \overline{f}\, \divergence\mathbf{V} + \overline{\mathbf{V}} \cdot \grad f, \vspace{-2mm}
 \label{eq:ODivergence1}
\end{equation}
where $(\bar{.})$ is the averaging process with an observationability at a length scale $\alpha$. \vspace{-1mm}
\end{definition}
As it will be clear shortly, this quantity represents the average flux of the vector field $\mathbf{F}$ out of a {\it finite} region with a representative length scale $\alpha$ defined by the averaging process $(\bar{.})$. For instance, for the Helmholtz filter 
 $f = \overline{f} - \alpha^2 \Delta \overline{f}$
 \label{eq:Helmholtz1}
where the representative scale of the finite averaging region is $\alpha$. The observation scale $\alpha$ could be imagined to be of similar size to the mesh size of the numerical simulation or the size of the measurement region (e.g., dimension of the wire in hotewire annemometry) in experiments. Note that, in a limit where the observable scale approaches zero this observable divergence reduces to the classical divergence defined in calculus. 

We now prove the main result governing the divergence theorem for an observable conservation law.\vspace{-1mm} 
\begin{theorem}
{{Observable Divergence Theorem.}} Let $\Omega$ be a region with surface boundary $S$ oriented outward. Let $\mathbf{F}(\mathbf{x}) = f \mathbf{V}(\mathbf{x}) =  (u(\mathbf{x}), v(\mathbf{x}), w(\mathbf{x}) ) f$ be a vector function in $C^1(\Omega)$. 
Then\vspace{-2mm}
\begin{equation*}
\hspace{-5mm}  \iiint_\Omega \!\! \odivergence\, \mathbf{F} \! = \!\!\! \iiint_\Omega \!\! \left( \overline{f}\, \divergence\mathbf{V} + \overline{\mathbf{V}} \cdot \grad f \right)  dV \!
  = \!\!\! \iint_S \!\! \mathbf{F \cdot n} \; dS\vspace{-2mm}
\end{equation*}
where $(\bar{.})$ is defined by the Helmholtz operator.\vspace{-1mm}
\end{theorem}

\begin{figure}
\begin{center}
  \includegraphics[angle=-90, width=.6\linewidth]{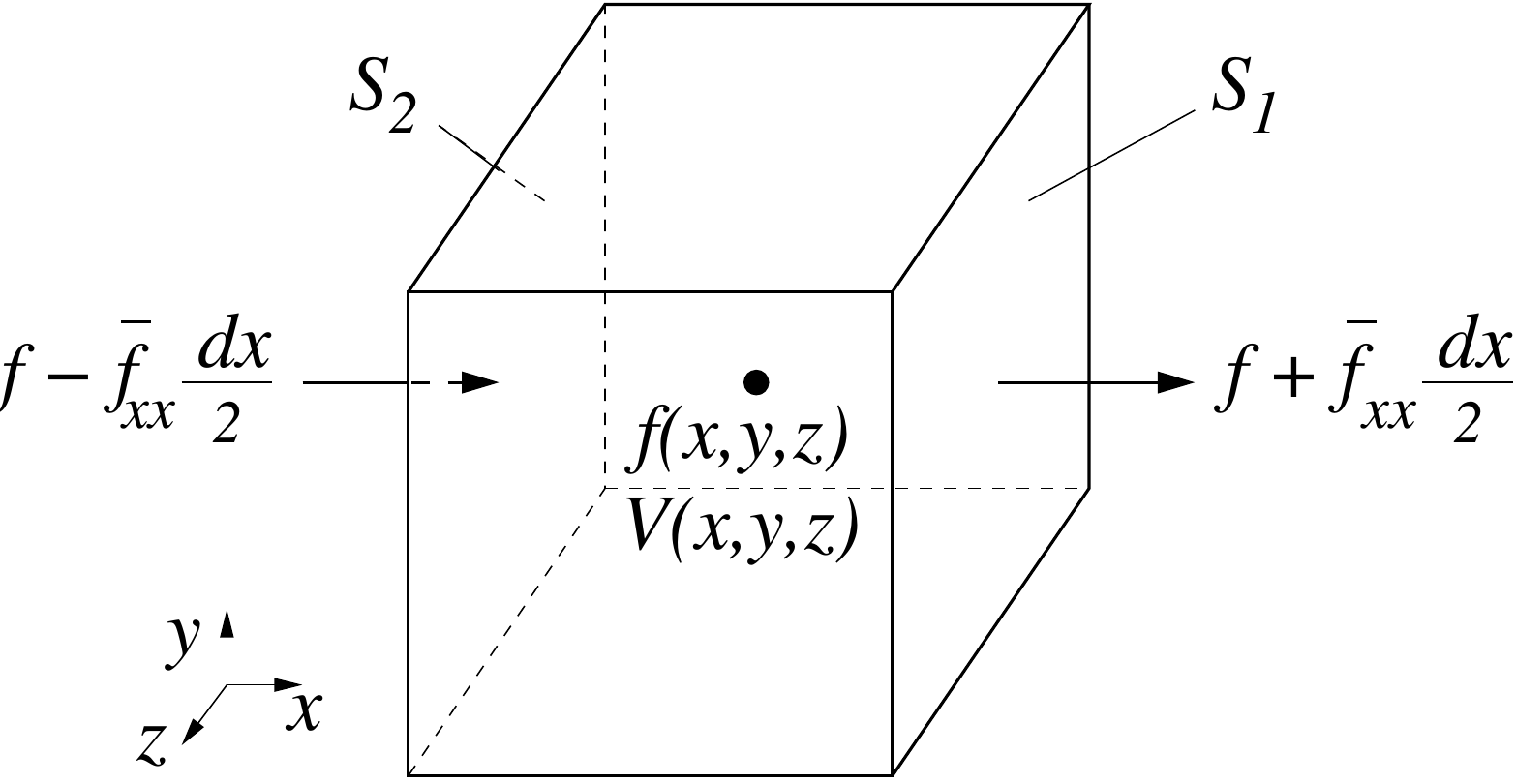}
\caption{A finite element of fluid at point $(x, y, z)$. }\vspace{-9mm}
\label{fig:FiniteElement1}
\end{center}
\end{figure}

{\noindent \bf Proof.}  
%
%
Consider a small closed element of fluid consisting of a rectangular parallelepiped centered around $(x, y, z)$, where the values of $f$ and $\mathbf{V}$ are defined, with edges parallel to the coordinate axes; see Figure \ref{fig:FiniteElement1}. 
If $f$ and $\bV$ are continuous the flux at the boundary point $\mathbf{x_s}$ of a finite volume is defined as $f(\mathbf{x_s}) \, \bV(\mathbf{x_s})$. 
%
%
%
We now start with the divergence definition in equation (\ref{eq:Divergence2}). The surface integral $\iint_S \mathbf{F \cdot n} \; dS$ over the rectangular parallelepiped can be regarded as a sum of six terms, one for each face of the parallelepiped. 
%
Now considering that $F_1 = f u$ and using the average flux approximation at the surface $S_1$ and $S_2$, one can write\vspace{-2mm}
\begin{align*}
 \dfrac{1}{\Delta V}  \iint\limits_{S_1+S_2} \mathbf{F \cdot n} dS & \simeq \dfrac{F_1( x+\frac{\Delta x}{2}, y, z) - F_1( x-\frac{\Delta x}{2}, y, z)}{\Delta x} \\[-2ex]
 & \hspace{-0mm} \simeq \dfrac{\overline{f} \overline{u} + \overline{f} u_x \frac{\Delta x}{2} + \overline{u} f_x \frac{\Delta x}{2}  + f_x u_x \frac{\Delta x^2}{4}} {\Delta x} \\
 & \hspace{-0mm} -  \dfrac{\overline{f} \overline{u} - \overline{f} u_x \frac{\Delta x}{2} - \overline{u} f_x \frac{\Delta x}{2} + f_x u_x \frac{\Delta x^2}{4} }{\Delta x} \\
 & = \overline{f} u_x + \overline{u} f_x + \mathcal{O}(\Delta x^2 )\vspace{-1mm}
\end{align*}
where we used the Taylor expansion and the definition of the Helmholtz filter to obtain an approximation for the flux on the face of the element\footnote{Since this is a rather important step here we reproduce the calculation of the values of a function at the edge of the parallelepiped based on the given values at the center of the volume. Using the Taylor expansion\vspace{-2mm}
\begin{equation}
  f(x+\frac{\Delta x}{2}) \simeq f(x) + f'(x) \dfrac{\Delta x}{2} + \mathcal{O}(\Delta x^2 )\vspace{-1mm}
\end{equation}
Using the Helmholtz equation we obtain\vspace{-2mm}
\begin{equation}
  f(x+\frac{\Delta x}{2}) \simeq \overline{f}(x) + f'(x) \dfrac{\Delta x}{2} + \mathcal{O}(\Delta x^2, \alpha^2 )\vspace{-2mm}
\end{equation}
Notice that in the above relation we have assumed that the averaging process is performed on a scale representative of the volume $\Delta V$. Therefore, the averaging length $\alpha$ only decreases up to an observable scale. Beyond that we are not able to resolve smaller scales. At that point, while the observable scale remains constant, $\Delta x$ could independently approach a mathematically infinitesimal value as required in calculus.}.
Adding the contributions from the rest of the faces for the parallelepiped and taking the limit of the volume $\Delta V$ approaching zero, one can write\vspace{-3mm}
\begin{align*}
 \lim_{\Delta V \rightarrow 0} & \dfrac{1}{\Delta V} \iint_{S} \mathbf{F \cdot n} dS  \\
  & = \overline{f} u_x + \overline{u} f_x + \overline{f} v_y + \overline{v} f_y + \overline{f} w_z + \overline{w} f_z \\
  & = \overline{f} \, \nabla \cdot \bV + \overline{\bV} \cdot \nabla f = \odivergence \, \mathbf{F}
\end{align*}
or\vspace{-8.8mm}
\begin{equation*}
  \iint_S \mathbf{F \cdot n} \; dS = \iiint_\Omega \odivergence\, \mathbf{F}.  \hspace{2cm}\mbox{} \hfill \blacksquare\vspace{-2mm}
\end{equation*}
One should note that the limit of vanishing volume is performed independently and beyond what is observable with the averaging scale $\alpha$. That is the volume of the element continuously shrinks until it is not observable at the available numerical or experimental resultion. At such a limit there is a clear distinction between $\odivergence\,  \mathbf{F}$ and $\divergence\, \mathbf{F}$. At the mathematical limit where both the volume element $\Delta V$ and the length scale of the averaging kernel $\alpha$ approch zero independently, the difference between the $\divergence\, \mathbf{F}$ and $\odivergence\, \mathbf{F}$ diminishes.

\paragraph{Observable Conservation Laws. \label{sec:OEquation}}
In this section we will develop the differential equations that must be satisfied by a fluid with an observability limit of $\alpha$. The equations are expected to resolve all relevant dynamical quantities up to this {\it observable} scale $\alpha$. As mentioned before, this scale is dictated by the resolution of a particular numerical simulation or the length scale resolution of a particular experimental equipment. For scales below $\alpha$ we assume that the medium still acts as a continuum and one can take the mathematical limit of $\Delta V \rightarrow 0.$ However, our {\it observable} scale remains at $\alpha$ even as this mathematical limit is taken to zero.  

Before the differential form of the  observable conservation laws for mass, momentum, and energy of a continuum are presented, we will formulate the general form of the observable differential conservation law for a quantity per unit volume, $f$. We assume $f$ satisfies a conservation law of the form\vspace{-2mm}
\begin{equation}
 \textrm{Rate of generation of $f$} = Q\vspace{-2mm}
\end{equation}
where $Q$ is the sum of all sources of $f$ inside the volume. For instance $Q$ is the total external force exerted on a control volume if one considers the conservation of momentum inside the volume. The rate of generation of $f$ is basically the outflow of $f$ minus the inflow at the boundaries of the volume plus the storage inside the volume. The rate of outflow minus the inflow at the boundaries was conveniently calculated in the previous section by the operator $\odivergence \,f \bV$ where $\bV$ is the continuum velocity. Therefore, the observable conservation of $f$ reduces to the following differential equation\vspace{-2mm}
\begin{equation}
  \dfrac{\p f}{\p t} + \overline{f}\ \divergence \bV + \overline{\bV} \cdot \grad f = Q.\vspace{-2mm}
  \label{eq:Of1}
\end{equation}
We now show that this equation satisfies the conservation of $f$ over a volume.\vspace{-2mm}

\begin{theorem}
{{Conservation of Conserved Quantities.}} Consider the observable evolution equation for $f$\vspace{-2mm}
\begin{equation}
  \dfrac{\p f}{\p t} + \overline{f}\ \divergence \bV + \overline{\bV} \cdot \grad f = 0.\vspace{-2mm}
  \label{eq:Of2}
\end{equation}
where $\overline{f}=g*f$ is the averaged or filtered quantity, and $g$ is an averaging kernel satisfying the properties in Table \ref{FilterProperties}. It can be shown that the observable evolution equation for $f$ conserves the total of $f$ over the whole domain; that is\vspace{-5mm}
\begin{equation}
  \frac{\p}{\p t}\int f(\mathbf{x})\, d\mathbf{x}=0.\vspace{-2mm}
\end{equation}
\end{theorem}

{\noindent \bf Proof.} Integrating the equation (\ref{eq:Of2}) over the domain and using the definition of the averaging, one obtains\vspace{-2mm}
\begin{align}
  \frac{\p}{\p t}\int f(\mathbf{x}) & \, d\mathbf{x}  = \iiint \left(  \nabla \cdot \mathbf{V}(\mathbf{x}) \iiint f(\mathbf{y}) g(\mathbf{x}-\mathbf{y}) d\mathbf{y}   \right. \nonumber \\
 &  \left. + \nabla f(\mathbf{x}) \cdot \iiint \mathbf{V}(\mathbf{y}) g(\mathbf{x}-\mathbf{y}) d\mathbf{y} \right) d\mathbf{x}.
  \label{eq:200}
\end{align}
Using the identities $f\nabla \cdot \mathbf{V} + \mathbf{V}\cdot \nabla f = \nabla \cdot (f\mathbf{V}) $ and $\iiint \left( f \nabla \mathbf{V} + \mathbf{V}\cdot \nabla f \right) d\mathbf{x}=0$ for functions that rapidly decay toward the boundary one can easily change the order of integration to prove that the right hand side of equation (\ref{eq:200}) is zero. 
\mbox{} \hfill $\blacksquare$

\paragraph{Observable Euler Equations. \label{sec:OEuler}}
Now, the observable compressible Euler equations can be easily derived by using the conservation of mass ($\rho$), momentum ($\rho \mathbf{V}$), and energy ($\rho e_T = \rho(e + \frac{1}{2} V^2)$) to obtain\vspace{-3mm}
\begin{subequations}\label{eq:OEuler1}
\begin{align}
  \dfrac{\p \rho}{\p t} + \overline{\rho}\ \divergence \bV + \overline{\bV} \cdot \grad \rho &= 0.
  \label{eq:OEulerMass1} \\
  \dfrac{\p \rho \bV}{\p t} + \overline{\rho\bV} \ \divergence \bV + \overline \bV \cdot \grad (\rho \bV) &= -\nabla p.
  \label{eq:OEulerMom1}\\
  \dfrac{\p \rho e_T}{\p t} + \overline{ \rho e_T} \ \divergence \bV + \overline{\bV} \cdot \grad (\rho e_T) & =  \nonumber \\
  & \hspace{-44mm}- \left( \overline{p}\ \divergence \bV + \overline{\bV} \cdot \grad p \right) + \rho \bV \cdot \mathbf{\Sigma} + S,
  \label{eq:OEulerEn1}
\end{align}
\end{subequations}
where $\mathbf{\Sigma}$ is the sum of all external forces acting on the control volume or its surface and $S$ is the sum of all energy sources inside the control volume. Note that in the limit of $\alpha$ approaching zero, that is when the observable scale approaches a mathematical zero, one recovers the classical Euler equations. 

In our previous publications \cite{Mohseni:10k}, we presented some mathematical and numerical results for the existence and uniqueness of the solution in 1D Euler along with the convergence to the entropy solution when $\alpha \rightarrow 0$. We also showed that our 1D observable Euler equations support the same traveling shock solution as the Euler equations.

\paragraph{Observable Incompressible Euler Equations.}
For the case of constant density, the observable continuity equation (\ref{eq:OEulerMass1}) reduces to the classical case of a divergence free velocity field; that is
$  \divergence \bV = 0.$
In this case, the observable momentum equation (\ref{eq:OEulerMom1}) can be simplified to\vspace{-2mm} 
\begin{equation}
  \dfrac{\p \bV}{\p t} + \overline{\bV} \cdot \grad \bV = -\frac{1}{\rho}\nabla p.\vspace{-2mm}
\end{equation}
This is exactly the inviscid Leray equation\cite{LerayJ:34a}, introduced without a formal derivation in 1934. A formal derivation of these equations from basic principles is offered here.

\paragraph{Numerical Experiments.} Several numerical experiments with increasing degrees of complexity were performed in order to numerically evaluate the performance of the observable Euler equations. 
These include benchmark test cases from \cite{Moin:10a} using the Euler/Navier-Stokes equations; 1D Sod, 1D Shu-Osher, 2D shock-vorticity/entropy wave interaction, 3D Taylor-Green vortex, 3D isotropic compressible turbulence, and 3D shock-turbulence interaction. See \cite{Mohseni:10p, Mohseni:10k} for the details of our numerical experiments. In order to reduce potential numerical dissipation a pseudo-spectral discretization technique is used. The equations are advanced in time with a Runge-Kutta-Fehlberg predictor/corrector (RK45). 
Spatial derivatives and the inversion of the Helmholtz operator were computed in the Fourier domain.  The terms were converted into the Fourier domain using a Fast Fourier Transform, multiplied by the appropriate term and then converted back into the physical domain.

\begin{figure}[!h]
\begin{center}
\vspace{-3mm}
\begin{minipage}{0.49\linewidth}\begin{center}
  \includegraphics[width=\linewidth]{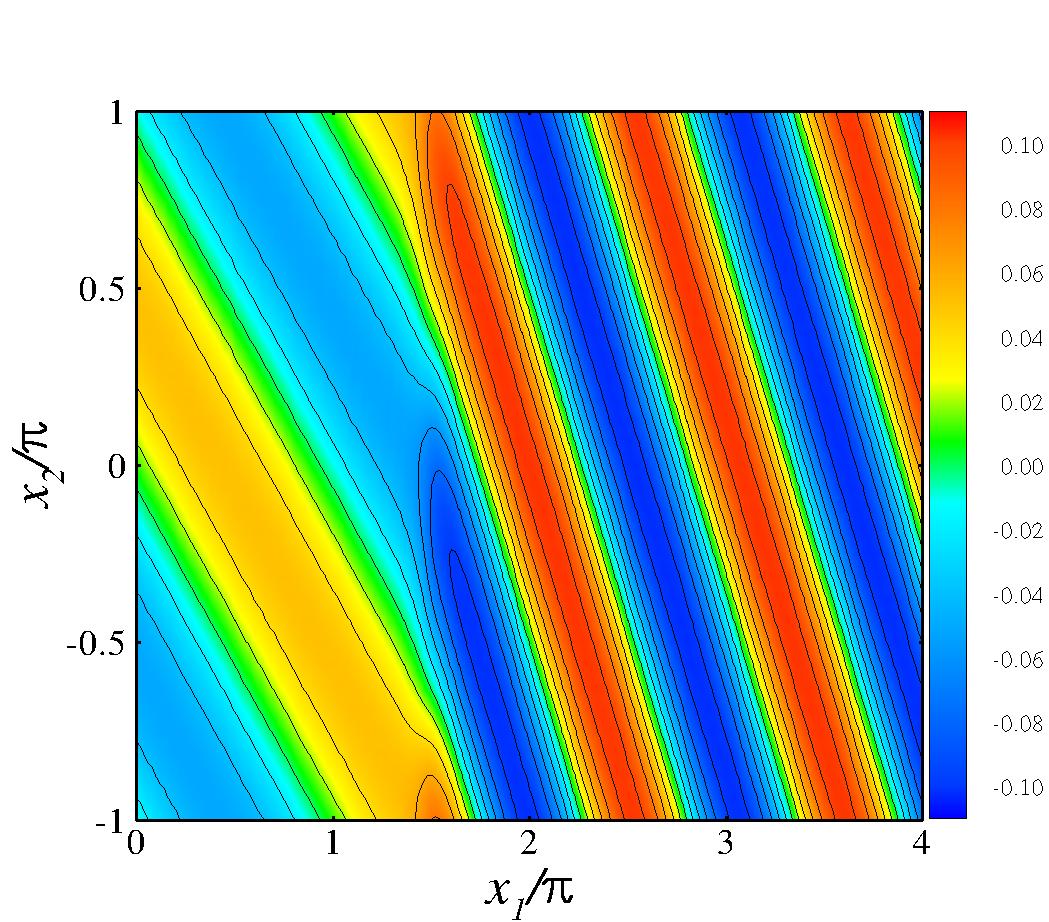}
\end{center}\end{minipage}
\begin{minipage}{0.49\linewidth}\begin{center}
  \includegraphics[width=\linewidth]{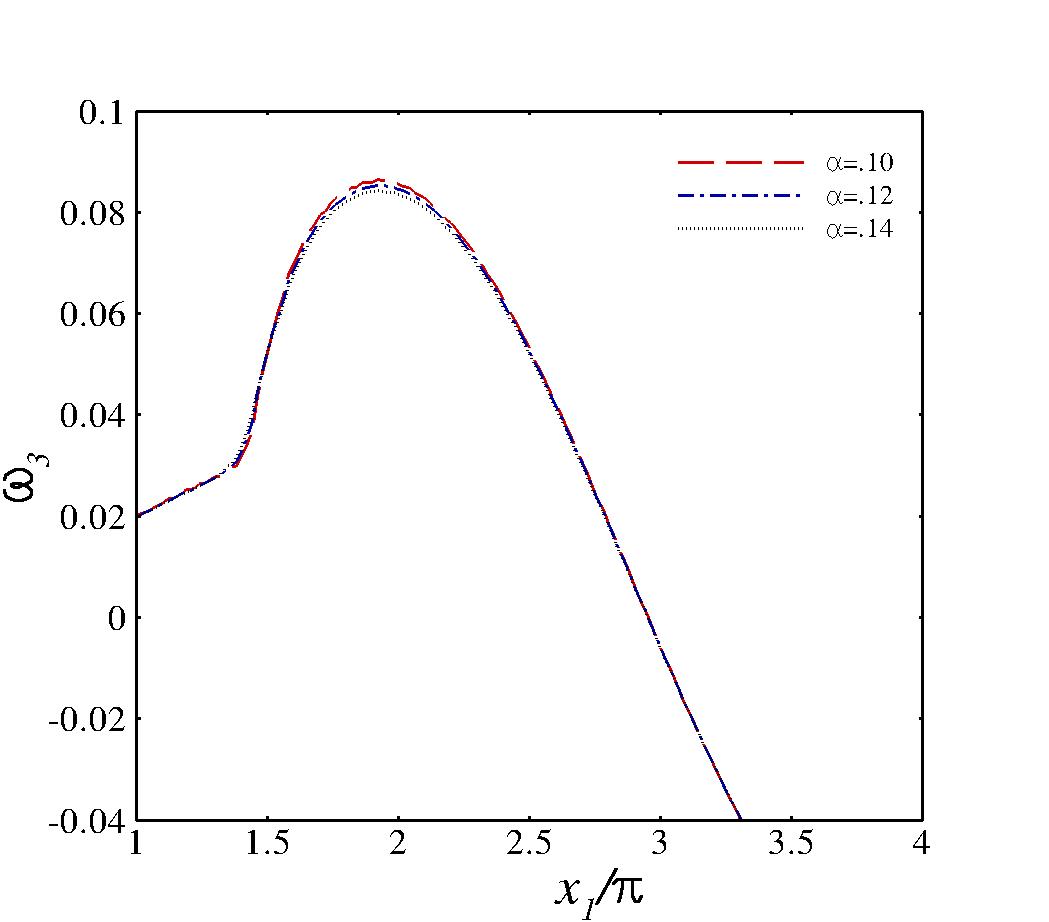}
\end{center}\end{minipage}\vspace{-8mm}
\end{center}
\caption{\label{Figure1}
\small Pseudo spectral simulation of the observable Euler equation. (Left) Vorticity contours for $\phi=45^{o}$ at $t=25$ and $k_1=k_2=1$. (Right) Vorticity Profiles at $x_{2}=0$, t=32, $\phi=75^{o}$, and $k_1=1$ where $\phi$ is the angle of the incoming wave and $k_1$ is wavelength of the incoming waves. \vspace{-5mm}
}
\end{figure}

Here we only present the results for the 2D interaction of a vorticity/entropy wave with a normal shock located at $x=1.5 \pi$. Identical initial and boundary conditions and similar spatial resolutions as in \cite{Moin:10a} is used here. Figure \ref{Figure1} shows the instantaneous vorticity contours and profiles for two different incident angles. The inflow vorticity/entropy wave interacts with the shock and changes its propagation direction and wavelength. In Figure \ref{Figure1}(b) instantaneous vorticity profiles for different $\alpha$ are shown for the incident angle $\phi=75^{o}$. Compared to the solutions in \cite{Moin:10a}, the vorticity profiles based on observable Euler equations do not show any spurious oscillations across the shock. Again note that this is done {\it without any viscous terms}. 
These results demonstrates the ability of the observable Euler equations to simulate 2D flows with shocks and resolve the shock  without any viscous terms or side effects (numerical dissipation) of using one-sided schemes. Also note that the thickness of the shock is controlled by the parameter $\alpha$, the width of the averaging Krenel.  For smaller $\alpha$, higher resolutions are required. 

We expect to employ the observable divergence theorem in future to other field quantities in order to obtain  regularized and observable evolution equations in other areas beyond the fluid dynamic problems considered here.\vspace{-1mm}

\mbox{}\vspace{-8mm} \bibliography{../../../RefA1}

\end{document}